\documentclass{kluwer}    

\newdisplay{guess}{Conjecture}
\usepackage{graphicx}
\begin{document}                                                                                   
\begin{article}
\begin{opening}         
\title{Evolutionary Synthesis Models for Galaxy Transformation in Clusters} 
\author{Jens \surname{Bicker}, U. \surname{Fritze-v. Alvensleben},  K.~J. \surname{Fricke}}  
\runningauthor{Bicker et al.}
\runningtitle{Evolutionary Synthesis Models for Galaxy Transformation in Clusters}
\institute{Universit\"ats-Sternwarte G\"ottingen}
\date{30.09.2002}

\begin{abstract}
The galaxy population in rich local galaxy clusters shows a ratio of one quarter
elliptical galaxies, two quarters S0 galaxies, and one quarter spiral galaxies.
Observations of clusters at redshift 0.5 show a perspicuously different ratio, the
dominant galaxy type are spiral galaxies with a fraction of two quarters while the
number of S0 galaxies decreases to a fraction of one quarter (Dressler et al.
1997). This shows an evolution of the galaxy population in clusters since redshift
0.5 and it has been suspected that galaxy transformation processes during the
infall into a cluster are responsible for this change. These could be merging,
starburst or ram-pressure stripping. We use our evolutionary synthesis models to
describe various possible effects of those interactions on the star formation of
spiral galaxies infalling into clusters. We study the effects of starbursts of
various strengths as well as of the truncation of star formation at various epochs
on the color and luminosity evolution of model galaxies of various spectral types.
As a first application we present the comparison of our models with observed
properties of the local S0 galaxy population to constrain possible S0 formation
mechanisms in clusters. Application to other types of galaxies is planned for the
future.
\end{abstract}
\end{opening}           

\section{Evolutionary Synthesis Model}
We use our evolutionary synthesis models based on Tinsley's equations and
following the stellar population through the HR diagram (for details see Fritze-v.
Alvensleben, Gerhard  1994) using stellar evolution tracks (Geneva), standard IMF
(Scalo), one metallicity (1/2 solar) and star formation rate (SFR) specific for
each galaxy type (Sandage 1986). This model well describes undisturbed galaxies,
in terms of average luminosity, colors (U...K), gas content and metallicity.  

To describe the galaxy interaction in clusters we use two scenarios: 
\begin{itemize}
\item Starburst for merging and tidal interaction of galaxies 
\item Star Formation Truncation for interaction with the Intra  Cluster Medium 
(ram-pressure-stripping)   
\end{itemize}
In case of the truncation we simply stop star formation at a given point in time.
A burst is described by a sudden increase of the SFR followed by an exponential
decline to a given value:
\begin{description}
\item[$\Psi_{\rm max}$: ]max. SFR at the beginning of the burst 
\item[$\tau_{\rm burst}$: ]decline time scale of the burst (typical $\sim 10^8$ yr) 
\item[$\Psi_{\rm f}$: ]the SFR after the burst (=0 or constant)  
\end{description}
The strength of the burst is defined by the fraction of gas which is
transformed into stars during the burst. Models produce the time evolution of 
magnitudes (Johnson UBVRIJHK) and colors for the different scenarios.

\section{Models}
\begin{figure}
\tabcapfont
\centerline{%
\begin{tabular}{c@{\hspace{1pc}}c}
\includegraphics[width=5.5cm]{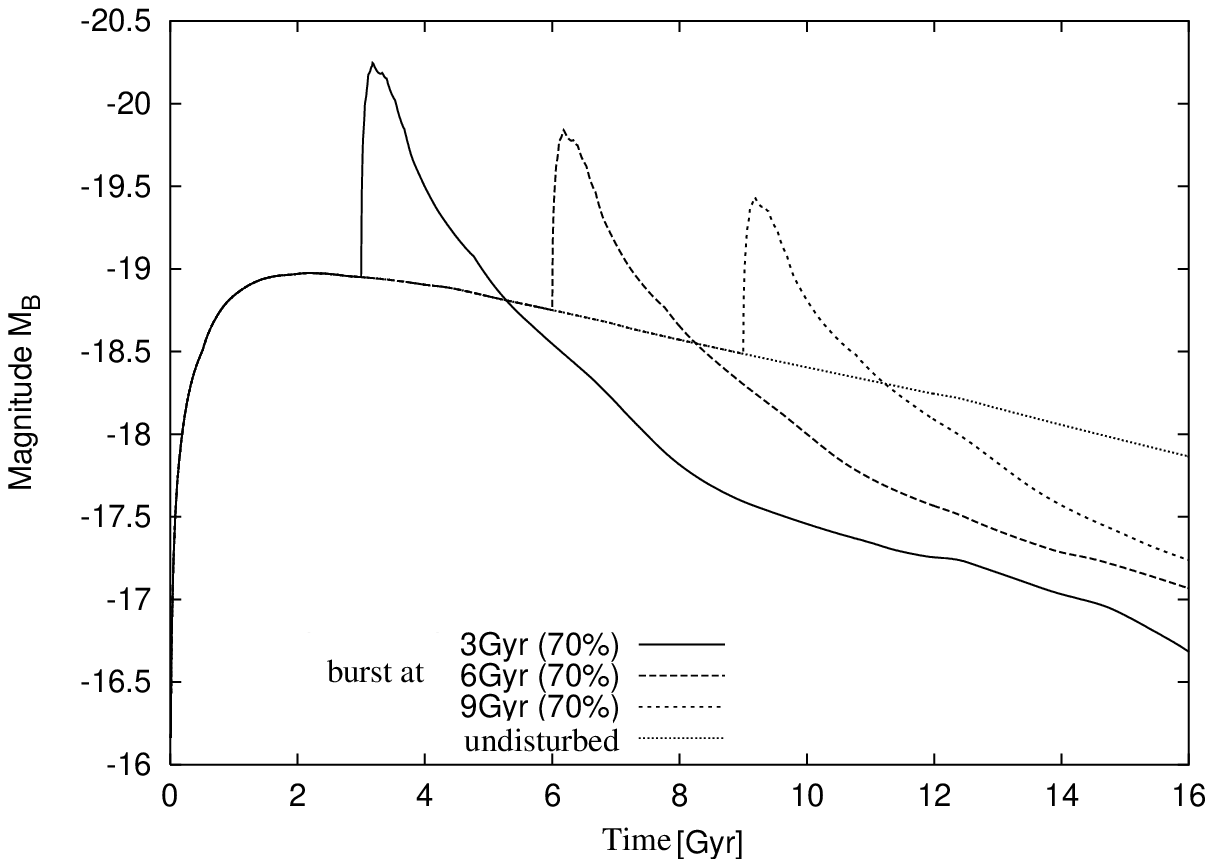} &
\includegraphics[width=5.5cm]{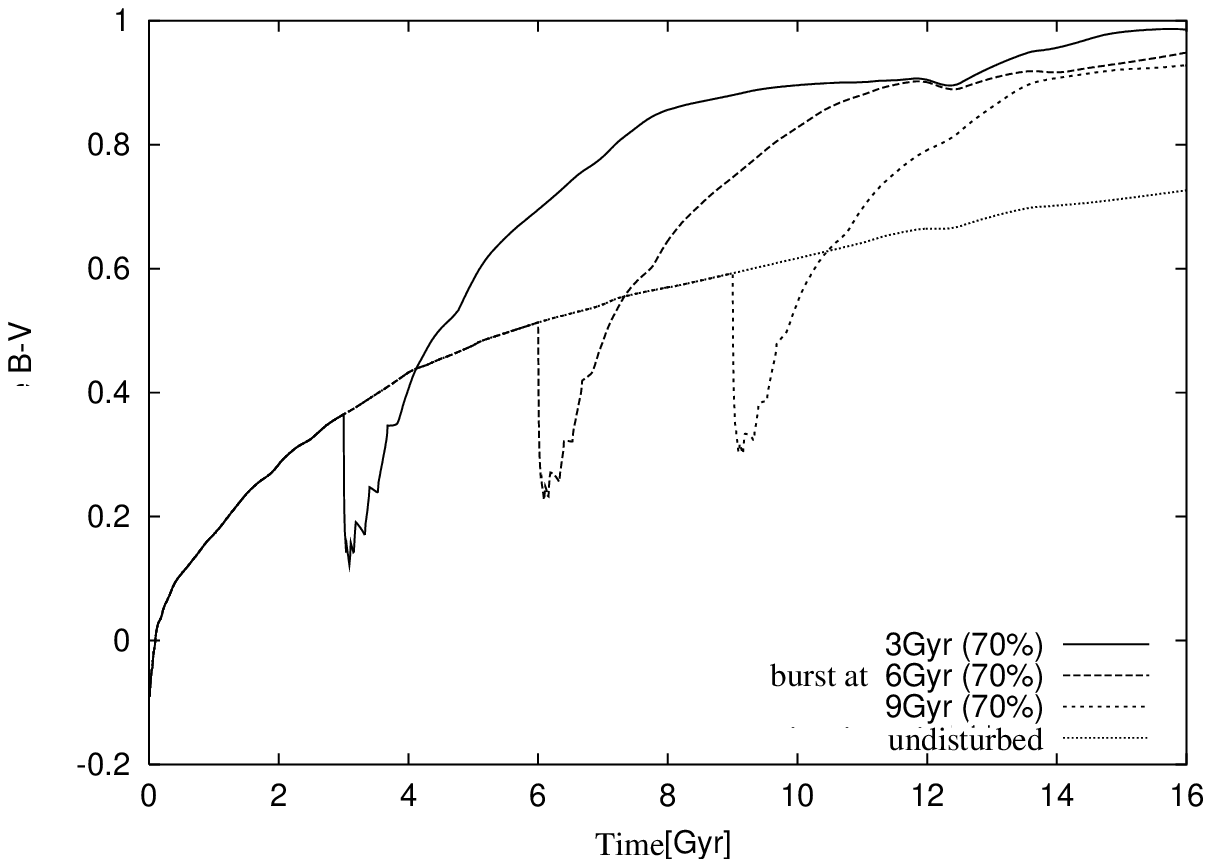} \\
a. & b. \\
\includegraphics[width=5.5cm]{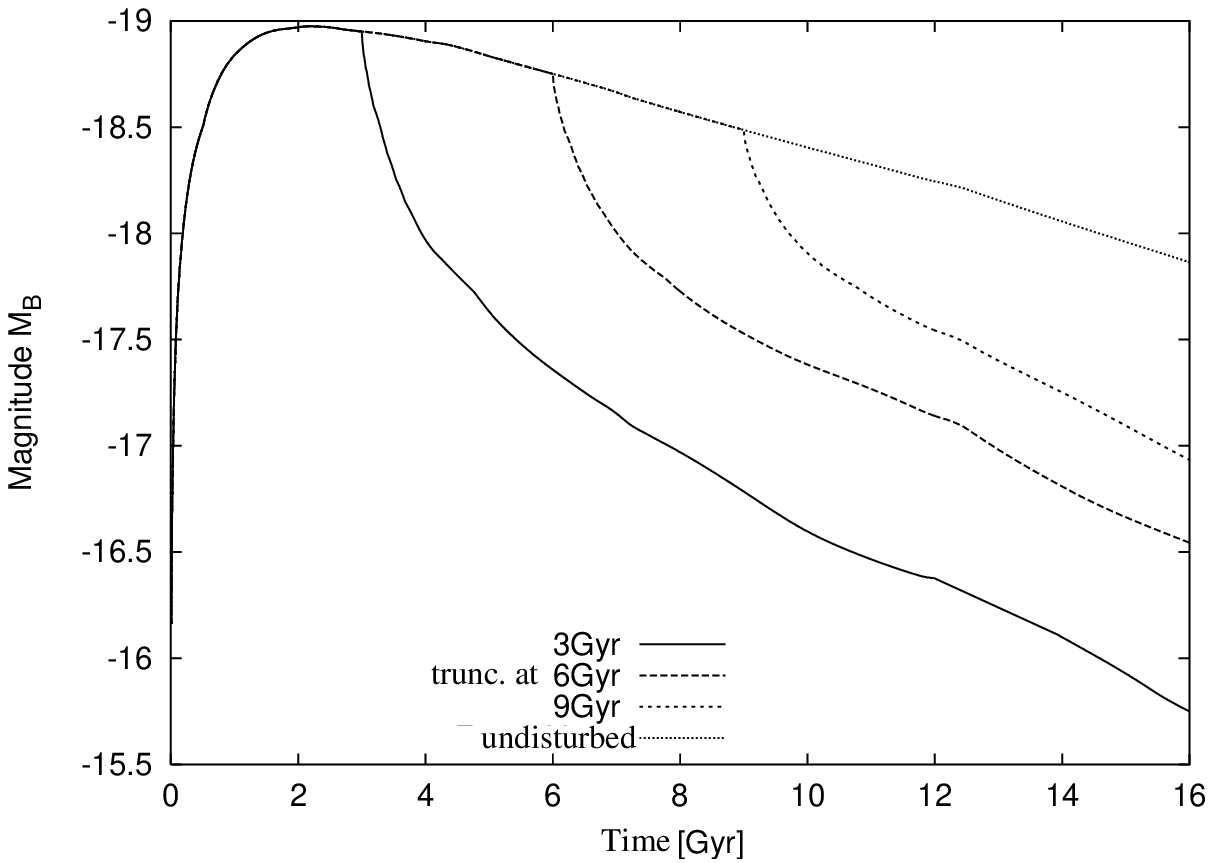} &
\includegraphics[width=5.5cm]{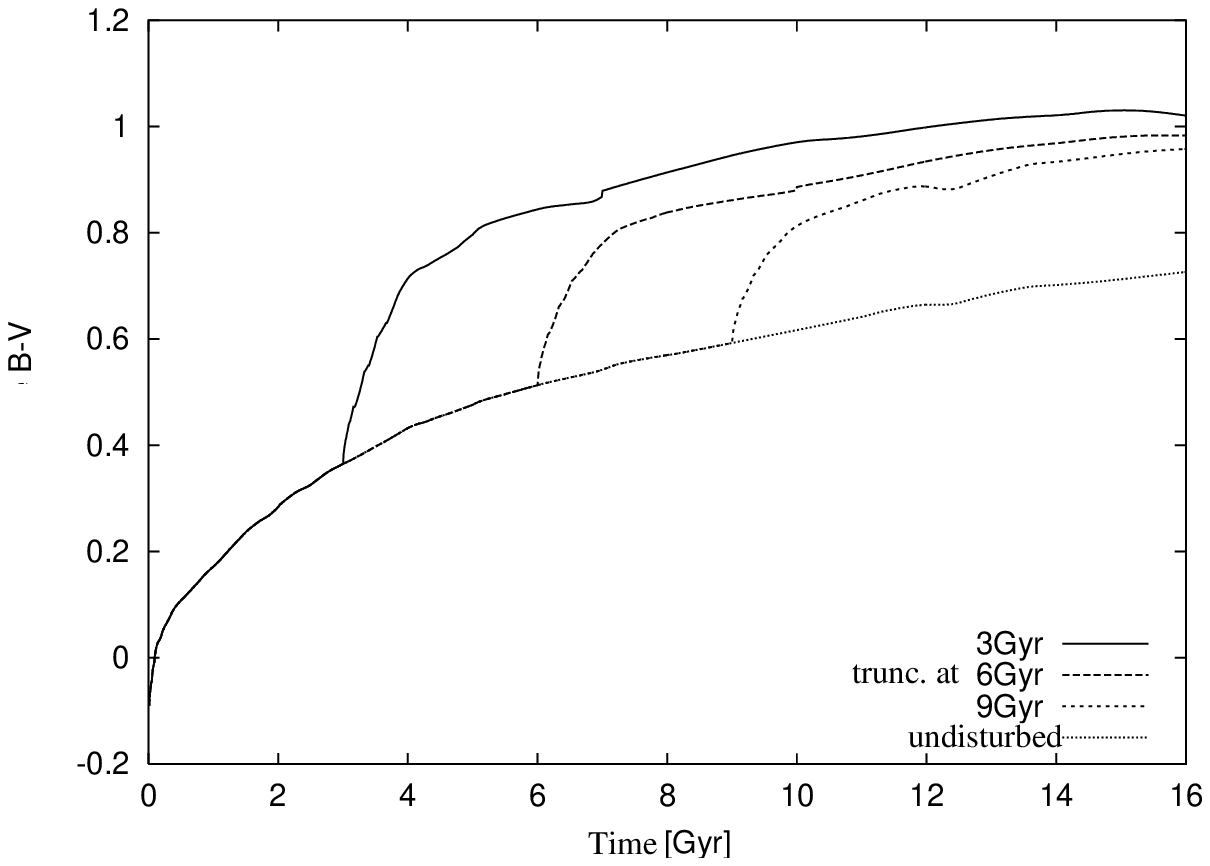} \\
c. & d. 
\end{tabular}}
\caption{Time evolution of ${\rm M_B}$ and ${\rm B-V}$ for Sb galaxies with strong bursts and no SF after it
(a. and b.) and for Sb galaxies with SF truncation (c. and d.)}
\label{m1}
\end{figure}

We compute models for a grid of parameters:
\begin{itemize}
\item Galaxy types Sa, Sb, Sc and Sd
\item Interaction (burst or  SF truncation) after 3, 6 or 9 Gyr of undisturbed evolution
\item Bursts transform into stars 70\%, 50\% or 30\% of the remaining gas
\item SFR after the burst of $\Psi_{\rm f}=1.5 M_\odot yr^{-1}$ or 0
\end{itemize}
Fig. \ref{m1} show an example of the time evolution of ${\rm M_B}$ and ${\rm B-V}$
for Sb galaxies with strong (70\%) bursts with $\Psi_{\rm f}=0$ at 3, 6 and 9 Gyr
of evolution (a,b) and for Sb galaxies with SF  truncation at the same points in
time (c,d). During the bursts the galaxies become  more luminous and bluer. But
after $\sim 2$ Gyr the luminosity falls below the luminosity  of an undisturbed Sb
galaxy, and ${\rm B-V}$ becomes redder. In case of SF truncation the luminosity fades
immediately and ${\rm B-V}$ becomes redder. It is mentionable that the burst
models as well as the truncation models reach nearly the same color in ${\rm B-V}$
at 4 Gyr after the interaction.

\section{S0 Galaxies}
As a first application of our models we look at S0 galaxies in clusters. As
mentioned in the Abstract, the S0 population evolves significantly from higher
redshifts to today.  
\begin{figure}
\tabcapfont
\centerline{%
\begin{tabular}{c@{\hspace{1pc}}c}
\includegraphics[width=5.5cm]{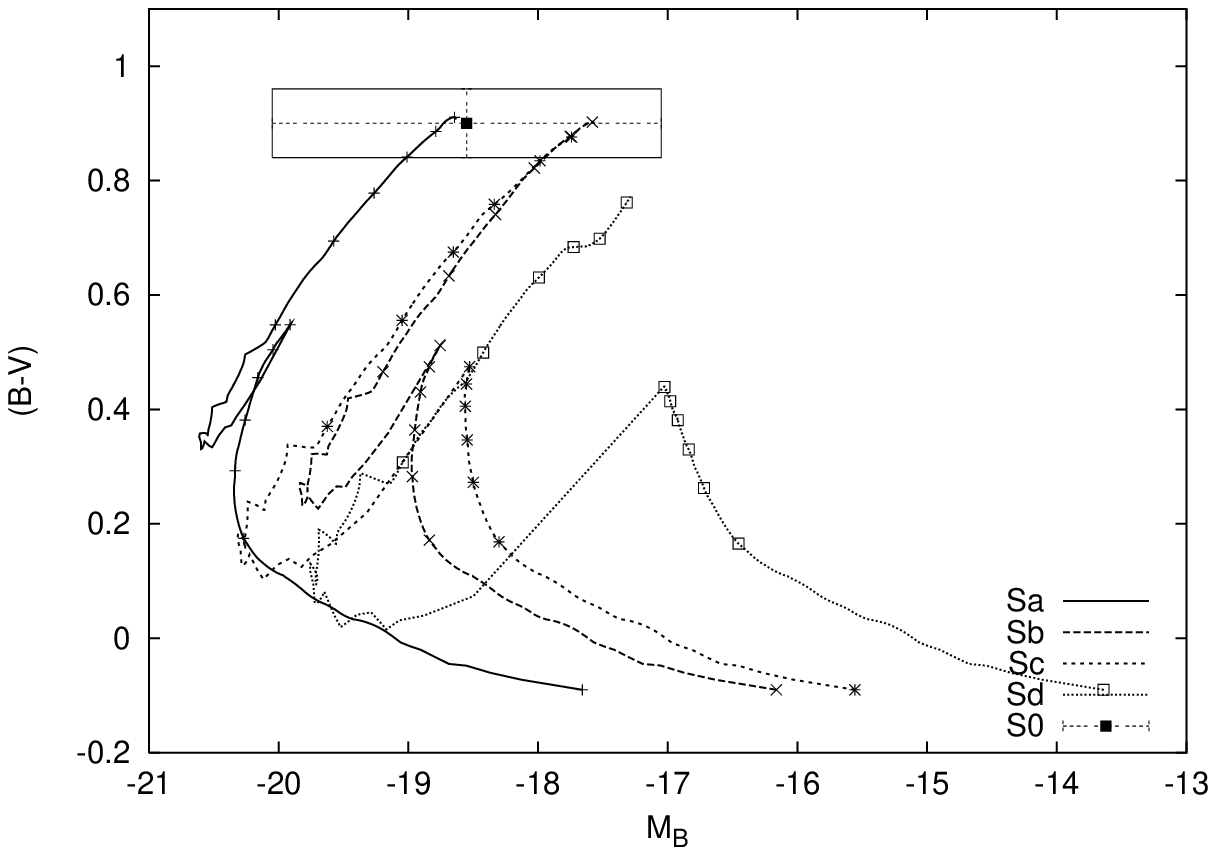} &
\includegraphics[width=5.5cm]{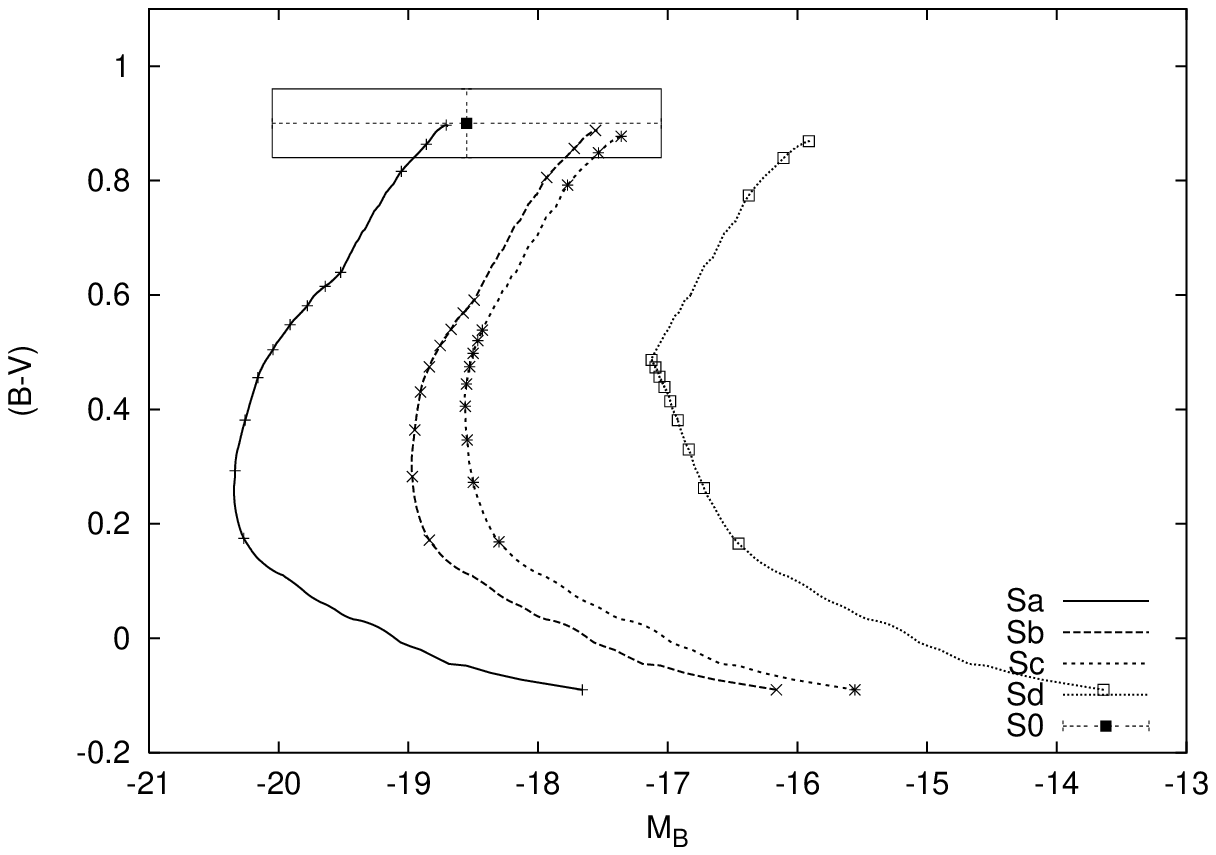} \\
a.~~ burst & b.~~ truncation
\end{tabular}}
\caption{Color-magnitude diagrams for the evolution of spiral galaxies
  with a.) strong bursts starting at an age of 6 Gyr and b.)
  star formation truncation after 9 Gyr. The symbols mark
  timesteps of 1 Gyr.}
\label{s01}
\end{figure}

Fig. \ref{s01}a shows a color-magnitude diagram for spiral models with a burst at 6 Gyr
and $\Psi_{\rm f}=0$ after the burst. Spiral galaxies evolve from the lower right to upper
left, get brighter and bluer during the burst and fade and redden thereafter
towards the observed location of local S0 galaxies as marked by the 1â box. Most
tracks end at 12 Gyr (today) in the region of S0 galaxies. Only the Sd galaxies
are too blue and slightly too faint.  If the burst model is to describe a merger
of equal type galaxies, (the mass of the bursting galaxy and) its luminosity is
to be increased by a factor 2 (-0.75 in magnitudes). The colors do not change.
(The symbols mark time steps of 1 Gyr)

Fig. \ref{s01}b shows spiral models with  SF truncation after 9 Gyr. At first the
galaxies evolve like in Fig. \ref{s01}a. After 9 Gyr the star formation stops and
the galaxies get redder and fade. Except for the Sd models all galaxies reach the
region of S0 galaxies.

For a detailed study see Bicker et al. 2002.

\section{Conclusion} 
Detailed comparison of photometric evolution models with local cluster S0
properties shows that most of the spirals falling into the cluster over an Hubble
time can be transformed into S0 galaxies (Bicker et al. 02): 
\begin{itemize}
\item The star formation must be stopped after the interaction (truncation or
burst with $\Psi_{\rm f}=0$) to reach the red  colors of S0 galaxies. 
\item Sd galaxies alone are too faint or too blue. Only in mergers the luminosity
can reach the S0 range. 
\item SF truncation may occur after $\ge 6$ Gyr, otherwise the galaxies would become
too red by today. 
\item Bursts must occur before $\sim 9$ Gyr. Thereafter the galaxies would become too
blue by today. 
\item Conclusions are the same for weak and strong bursts. 
\item Conclusions are the same for all colors (U...K). 
\end{itemize}

These results agree well with spectral analysis of  S0 galaxies (Jones et al.
2000). They also found that the progenitors of  S0 galaxies in rich clusters are
mostly early-type spirals that  had their star formation truncated in the cluster
environment.


\theendnotes

\end{article}

\begin{thebibliography}{}

\bibitem[\protect\citeauthoryear{Bicker}{2002}]{Bicker}
Bicker, J., Fritze-v. Alvensleben, U., Fricke, K.~J.
\newblock {\em A\&A}, 387:421, 2002.

\bibitem[\protect\citeauthoryear{dressler}{1997}]{Dressler}
Dressler, A., Oemler A.~J., Couch, W.~J. et al.
\newblock {\em AJ}, 490:577, 1997.

\bibitem[\protect\citeauthoryear{Fritze}{1994}]{Fritze}
Fritze-v. Alvensleben U., Gerhard O. E.
\newblock {\em A\&A}, 185:751, 1994.

\bibitem[\protect\citeauthoryear{jones}{2000}]{jones}
Jones, L., Smail, I., Couch, W.~J. 
\newblock {\em AJ}, 528:118, 2000.

\bibitem[\protect\citeauthoryear{sandage}{1986}]{sandage}
Sandage, A.
\newblock {\em A\&A}, 161:89, 1996.

\end{thebibliography}
\end{document}